%% file: main.tex
\documentclass[twocolumn]{article}

\usepackage{PRIMEarxiv}

\usepackage{authblk}

\input{utils/include}

\input{utils/general_utils}

\input{utils/math_utils}

\usepackage[utf8]{inputenc} % allow utf-8 input
\usepackage[T1]{fontenc}    % use 8-bit T1 fonts
\usepackage{hyperref}       % hyperlinks
\usepackage{url}            % simple URL typesetting
\usepackage{booktabs}       % professional-quality tables
\usepackage{amsfonts}       % blackboard math symbols
\usepackage{nicefrac}       % compact symbols for 1/2, etc.
\usepackage{microtype}      % microtypography
\usepackage{lipsum}
\usepackage{fancyhdr}       % header
\usepackage{graphicx}       % graphics
\usepackage{float}

\begin{document}

\twocolumn[\begin{@twocolumnfalse}
\input{content/head}
\input{content/abstract} 

\end{@twocolumnfalse}]

\input{content/content}

\FloatBarrier

%Bibliography
\bibliographystyle{unsrt}  
% \bibliography{references}  

% \newpage
% \appendix
% \onecolumn
% \input{content/supplementary}

\end{document}

%% file: utils/include.tex
\usepackage{microtype}
\usepackage{graphicx}
\usepackage{subfigure}
\usepackage{booktabs} % for professional tables

\usepackage{amsmath}
\usepackage{amssymb}
\usepackage{mathtools}
\usepackage{amsthm}
\usepackage{MnSymbol}
\usepackage{wasysym}

\usepackage{hyperref}
\usepackage[capitalize,noabbrev]{cleveref}

\usepackage[textsize=tiny]{todonotes}

\usepackage{multirow} 
\usepackage{pifont}
\usepackage{color, colortbl}

\usepackage{blindtext}
\usepackage{lipsum}
\usepackage{threeparttable}

\usepackage{wrapfig}
\usepackage{times}
\usepackage{multirow}
\usepackage{url}

\usepackage{color}
\usepackage{xcolor}

\usepackage{enumitem}
\usepackage{xspace}
\usepackage{xfrac}
\usepackage{comment}
\usepackage{tikz}
\usepackage{etoolbox}

\usepackage{placeins}
\usepackage{subcaption}

\definecolor{PL_color}{rgb}{0.858, 0.188, 0.478}

\usepackage{pifont}
\usepackage{adjustbox}
\usepackage{threeparttable}
\usepackage{multirow}

\DeclareMathAlphabet\mathbfcal{OMS}{cmsy}{b}{n}

\DeclarePairedDelimiterX{\inp}[2]{\langle}{\rangle}{#1, #2}

\usepackage{acro}

\usepackage{color, colortbl}
\definecolor{Gray}{gray}{0.93}
\definecolor{Orange}{rgb}{1,0.5,0}
\definecolor{DGray}{gray}{0.83}
\definecolor{LightCyan}{rgb}{0.88,1,1}

\RequirePackage[normalem]{ulem} %DIF PREAMBLE
\RequirePackage{color}\definecolor{RED}{rgb}{1,0,0}\definecolor{BLUE}{rgb}{0,0,1} %DIF PREAMBLE
\usepackage{newfloat}
\DeclareFloatingEnvironment[name={Supplementary Figure}]{suppfigure}
\usepackage{sidecap}

\usepackage{tikz,xcolor}

\definecolor{cvprblue}{rgb}{0.21,0.49,0.74}

\usepackage{arydshln}

%% file: utils/general_utils.tex
\usepackage{pifont}

\usepackage{color, colortbl}
\definecolor{Gray}{gray}{0.93}
\definecolor{Orange}{rgb}{1,0.5,0}
\definecolor{DGray}{gray}{0.83}
\definecolor{LightCyan}{rgb}{0.88,1,1}

\usepackage[most]{tcolorbox}

%% file: utils/math_utils.tex
\usepackage{amsmath,amsfonts,bm}

\def\eqref#1{(\ref{#1})}
\def\1{\bm{1}}
\newcommand{\train}{\mathcal{D}}

\def\vx{{\bm{x}}}

\def\mA{{\bm{A}}}

\def\mF{{\bm{F}}}

\def\mJ{{\bm{J}}}

\DeclareMathAlphabet{\mathsfit}{\encodingdefault}{\sfdefault}{m}{sl}
\SetMathAlphabet{\mathsfit}{bold}{\encodingdefault}{\sfdefault}{bx}{n}

\def\gX{{\mathcal{X}}}
\def\gY{{\mathcal{Y}}}

\newcommand{\Ptrain}{\hat{P}_{\rm{data}}}

\newcommand{\Ptest}{\hat{P}_{\rm{test}}}
\newcommand{\R}{\mathbb{R}}

%% file: content/head.tex
\title{DECIPHERING THE DEFINITION OF ADVERSARIAL ROBUSTNESS FOR POST-HOC OOD DETECTORS
%%%% Cite as
%%%% Update your official citation here when published 
% \thanks{\textit{\underline{Citation}}: 
% \textbf{Authors. Title. Pages.... DOI:000000/11111.}} 
}

\author[1,2]{Peter Lorenz}
\author[1,3]{Mario Fernandez}
\author[2]{Jens Müller}
\author[2]{Ullrich Köthe}
\affil[1]{Fraunhofer ITWM, Germany}
\affil[2]{Heidelberg University, Germany}
\affil[3]{\'{E}cole Normale Sup\'{e}rieure, PSL University, France}

% \author{
%   Peter Lorenz \\
%   Heidelberg University \\
%   Univ \\
%   City\\
%   \texttt{\{Author1, Author2\}email@email} \\
%   %% examples of more authors
%    \And
%   Mario Fernandez \\
%   ITWM Fraunhofer \\
%   Univ \\
%   \AND
%   Jens Müller, Ullrich Köthe \\
%   Heidelberge University \\
% }

 \maketitle

%% file: content/abstract.tex
\begin{abstract}
Detecting out-of-distribution (OOD) inputs is critical for safely deploying deep learning models in real-world scenarios. 
In recent years, many OOD detectors have been developed, and even the benchmarking has been standardized, i.e. OpenOOD. 
The number of post-hoc detectors is growing fast. They are showing an option to protect a pre-trained classifier against natural distribution shifts and claim to be ready for real-world scenarios.
However, its effectiveness in dealing with adversarial examples (AdEx) has been neglected in most studies.
In cases where an OOD detector includes AdEx in its experiments, the lack of uniform parameters for AdEx makes it difficult to accurately evaluate the performance of the OOD detector.
The existing adversarial OOD definition only declares that AdEx is semantic preserving.
We extend this definition by including the attention maps of the underlying neural network classifier and develop a Grad-CAM-based metric to measure the attention shift on the distributional level.
This paper investigates the adversarial robustness of 16 post-hoc detectors against various evasion attacks.
It also discusses a roadmap for adversarial defense in OOD detectors that would help adversarial robustness. 
We believe that level 1 (AdEx on a unified dataset) should be added to any OOD detector to see the limitations. 
The last level in the roadmap (defense against adaptive attacks) we added for integrity from an adversarial machine learning (AML) point of view, which we do not believe is the ultimate goal for OOD detectors.

Code: \href{https://github.com/adverML/AdvOpenOOD}{github.com/adverML/AdvOpenOOD}

\keywords{adversarial examples \and OOD \and post-hoc detectors} 

\hspace{2cm}
\end{abstract}

% keywords can be removed
% \begin{IEEEkeywords}
% Adversarial examples, ViM, OOD, XAI
% \end{IEEEkeywords}

%% file: content/content.tex
\section{INTRODUCTION}
% https://arxiv.org/pdf/2110.11334

Adversarial robustness in the context of out-of-distribution (OOD) detection refers to the ability of a detector to correctly identify OOD samples even when they have been adversarial perturbed to evade a deep neural network (DNN).
Evasion attacks, e.g. PGD \cite{madry2017towards}, which are designed to fool deep learning classifiers, are difficult to spot as an outlier for OOD detectors.
In general, detecting adversarial examples (AdEx) is a challenging task and is almost as hard as classifying them \cite{tramer2022detecting}.
To prevent errors in real-world applications, it is crucial to detect OOD cases, not only for natural distribution shifts \cite{taori2020measuring} but also for AdEx \cite{sehwag2019analyzing} without degrading the generalizability of the underlying pre-trained classifier \cite{yang2021generalized}.

Current standardized benchmarks such as OpenOOD \cite{zhang2023openood} and RoboDepth \cite{kong2024robodepth} merely focus on natural distribution shifts and corruptions \cite{hendrycks2019benchmarking}. 
Especially OpenOOD aims to make a fair comparison across methods initially developed for anomaly detection \cite{habib2023towards}, model uncertainty \cite{ledda2023adversarial}, open-set recognition (OSR) \cite{shao2022open}, and OOD detection \cite{lee2018simple} but neglecting AdEx.
% The focus is to detect distribution shifts as outliers when the classifier is trained on dataset A, but evaluated on dataset B. 

OpenOOD benchmark suite \cite{yang2021generalized} evaluates methods on semantic shift (e.g., samples that are semantically different from the training data, representing truly novel or unseen concepts.) \cite{hendrycks2016baseline} and covariate shift (e.g. samples that come from a different distribution than the training data, but still belong to the same semantic categories) \cite{zhang2023openood}.
% Earlier in \cite{yang2021generalized}, distributional shifts have been discussed and can be caused by semantic shifts \cite{ben2010theory,li2017deeper,wang2018deep}.
\begin{table}[tb]
\centering
\caption{\small
Post-hoc OOD detectors architecture comparison: 
I) Features: output of layers before the last layer.
II) Logits: raw output of the last layer.
III) Probabilities: normalized output of the last layer.
IV) Adversarial robust against evasive attacks (see 
% \cref{sec:adv_defenses} 
section 2.3
and \cref{tab:results}).
The `$\thicksim$' means that the detector only partly fulfills a certain property. 
All methods are in the OpenOOD benchmark suite \cite{zhang2023openood}.
} \label{tab:taxonomy}
\resizebox{\columnwidth}{!}{%
\begin{tabular}{ll|ccc|c}
\toprule[1pt]
\hline
\multirow{2}{*}{\textbf{Detector}} & \multirow{2}{*}{\textbf{Venue}} & \multicolumn{3}{c|}{\textbf{Detector Architecture}} & \multicolumn{1}{l}{\textbf{Adversarial}} \\
                &            &\textbf{Features} & \textbf{Logits} & \textbf{Probs} & \textbf{Robust} \\ \hline
        SCALE   & ICLR'24    & $\checkmark$ & $\checkmark$ &              &              \\ 
        NNGUIDE & NeurIPS'23 & $\checkmark$ &              & $\checkmark$ &              \\
        GEN     & CVPR'23    &              &              & $\checkmark$ &              \\
        ASH     & ICLR'23    & $\checkmark$ &              &              &              \\
        DICE    & ECCV'22    & $\checkmark$ &              &              &              \\
        KNN     & ICML'22    & $\checkmark$ &              &              &              \\
        VIM     & CVPR'22    & $\checkmark$ & $\checkmark$ &              &              \\
        KLM     & ICML'22     &              &              & $\checkmark$ &              \\
        MLS     & ICML'22     &              & $\checkmark$ &              &              \\
        REACT   & NeurIPS'21 & $\checkmark$ &              &              &              \\
        RMDS    & ARXIV'21   & $\checkmark$ &              & $\checkmark$ & $\thicksim$  \\
        GRAM    & ICML'20    &              & $\checkmark$ &              &              \\
        EBO     & NeurIPS'20 & $\checkmark$ &              &              &              \\
        ODIN    & ICLR'18    &              &              & $\checkmark$ &              \\
        MDS     & NeurIPS'18 & $\checkmark$ &              &              & $\thicksim$  \\
        MSP     & ICLR'17    &              &              & $\checkmark$ &              \\
\hline
\bottomrule[1pt]
\end{tabular}
}
\end{table}
Current OOD detection methods, as some listed in \cref{tab:taxonomy}, achieve outstanding results on prominent OOD benchmarks, such as the OpenImage-O \cite{wang2022vim}, ImageNet-O \cite{hendrycks2021natural}, Texture \cite{cimpoi2014describing}, and iNaturalist \cite{huang2021mos, van2018inaturalist}.
OOD detection is a very quickly growing field due to the number of methods added to OpenOOD.
More specific post-hoc methods with their plug-and-play capabilities on pre-trained classifiers are more flexible and scalable compared to methods that require full retraining on new OOD data \cite{yang2022openood,cong2022sneakoscope}.
Simple post-hoc methods like KNN \cite{sun2022out} are highlighted maintaining good performance on toy-datasets (e.g. MNIST \cite{deng2012mnist}, CIFAR-10 or CIFAR-100 \cite{krizhevsky2009cifar}), and also show outstanding performance on the more realistic dataset like ImageNet \cite{deng2009imagenet} according to \cite{yang2022openood}.
These experiments do neglect the adversarial robustness of ``state-of-the-art'' (sota) detectors and the real-world capabilities \cite{cui2022out} are questionable as past studies had shown \cite{sehwag2019analyzing, song2020critical, chen2020robust, salehi2021unified}.
% why is it questionable?
OOD detectors should be able to reliably detect inputs that come from a different distribution than the training data, even if the shift is subtle or complex.
AdEx remains challenging because they share the same semantics as the training data, but aim to modify the classifier's output. 

In this study, we investigate the adversarial robustness of post-hoc ODD detectors. Our contributions can be summarized as follows:
\begin{itemize}[nosep]
    \item We revise the definition of adversarial OOD regarding post-hoc OOD methods to finally have a common understanding of robust adversarial OOD detection referring to \cref{sec:definition}.
    Finally, we extended the existing definition with attention maps (see \cref{eq:attention}).
    \item We examine 16 post-hoc OOD detectors by delving into their current ability to detect AdEx — an aspect that has been disregarded.
    \item We expand the OpenOOD framework with evasive attacks and provide adversarial OOD datasets to have standardized attack parameters.
    Moreover, we analyze the semantic shift between benign and AdEx by using Grad-CAM combined with a novel metric.
    %\href{https://github.com/adverML/AdvOpenOOD}{github.com/adverML/AdvOpenOOD}.
    % \link{Code}{https://anonymous.4open.science/r/anonym-A6B5/README}
\end{itemize}
% \begin{links}
    % - \href{https://anonymous.4open.science/r/anonym-A6B5}{Link to the code and dataset.}
%     \link{Datasets}{https://osf.io/6ywmh/?view_only=ff96d6d9151a48e09497366782495dd3}
%     % \link{Extended version}{https://aaai.org/example/extended-version}
% \end{links}

\section{RELATED WORK}
This section discusses existing white-box evasion attacks, the benefits of post-hoc detectors, existing methods for detecting AdEx, and finally Grad-CAM, a tool for understanding the neural networks' attention shift initiated by AdEx.

\subsection{Evasion Attacks Crafting Inliers} \label{sec:evasionattacks}
The objective of evasion attacks is to generate AdEx that will result in the misclassification of inputs at deep learning models \cite{biggio2013evasion}. 
It is possible to distinguish between two types of attack: black-box attacks \cite{zheng2023blackboxbench}, where the classifier is queried, and white-box attacks \cite{carlini2017towards}, where the network is under the attacker's complete control.
A white-box threat model is strictly stronger.
Evasion attacks try to find the smallest possible perturbation, often imperceptible to humans, to manipulate the model's decision boundaries. 
% They exploit models by finding the smallest possible perturbation, often imperceptible to humans, to manipulate the model's decision boundaries. 

More formally, for an input \(\vx\) with the ground-truth label \(y\), an adversarial example \(\vx'\) is crafted by adding small noise $\boldsymbol \delta$ to \(\vx\) such that the predictor model loss \(\mJ(\vx', y)\) is maximized. 
The \(L^p\) norm of the adversarial noise should be less than a specified value \(\epsilon\), i.e., \(\left \| \vx - \vx' \right \| \leq \epsilon\), e.g., $\epsilon=8/255$ \cite{croce2020robustbench}, to ensure that the image does not change semantically.
The attack method Fast Gradient Sign Method (FGSM) by  \cite{goodfellow2014explaining} maximizes the loss function in a single step by taking a step towards the sign of the gradient of $\mJ(\vx,y)$ w.r.t. to $\vx$: $\vx' = \vx + \epsilon \text{sign}(\nabla_{\vx} \mJ(\vx', y))$,
% \begin{equation} \label{eq:fgsm}
% {
%     \vx' = \vx + \epsilon \text{sign}(\nabla_{\vx} \mJ(\vx', y)),
% }
% \end{equation}
where the noise meets the $L^\infty$ norm bound $\epsilon$.
Furthermore, this approach can be applied iteratively, as shown by  \cite{kurakin2018adversarial}, using a reduced step size $\alpha$: 
$vx'_0 = \vx,\;   \vx'_{t+1} =  \vx'_t + \alpha \epsilon \text{sign}(\nabla_{\vx} \mJ(\vx'_t, y))$,
% \begin{equation} \label{eq:pgd}
% {
    % \vx'_0 = \vx,\;   \vx'_{t+1} =  \vx'_t + \alpha \epsilon \text{sign}(\nabla_{\vx} \mJ(\vx'_t, y)), 
% }
% \end{equation}
where in each step, the perturbation should be constrained within the $L^\infty$ ball of radius $\epsilon$.
This constraint is a characteristic of the Projected Gradient Descent (PGD) attack proposed by \cite{madry2017towards}, which is commonly considered a standard attack for the evaluation of model robustness \cite{liu2023comprehensive}.
{Masked PGD (mPGD)} \cite{xu2023patchzero} is a variant of the PGD attack that restricts perturbations to a specific region within an image. 
The PGD attack is extended towards a patch: $\vx_{t+1}' = \text{Clip}_{\vx'} \left( \vx_{t}' + \alpha \cdot \text{sign}(\nabla_{\vx} \mJ(\vx_{t}', y, \bm{\theta})[patch]), \epsilon \right).$
% \begin{equation} \label{eq:maskedpgd}
% {\scriptstyle
% \vx_{adv}^{(t+1)} = \text{Clip}_{\vx} \left( \vx_{{adv}}^{(t)} + \alpha \cdot \text{sign}(\nabla_{\vx} \mJ(\vx_{{adv}}^{(t)}, y, \bm{\theta})[patch]), \epsilon \right).
% }
% \end{equation}
In this context, the term ``patch'' refers to the region specified as $[x: x + h, y: y + w]$, where $[x, y, h, w]$ are the provided patch's coordinates and dimensions.
AutoAttack (AA) \cite{croce2020reliable} consists of four attacks, with the first two being variants of PGD. 
It remains uncertain whether AA is superior to PGD \cite{lorenz2022unfolding, lee2023robust, cina2024attackbench}. 
Recently, \cite{yang2023rethinking} modified PGD, which could also enhance the effectiveness of AA and show that PGD is an essential part of AA.

Lastly, DeepFool (DF) \cite{moosavi2016deepfool} assumes that the network's decision boundary is linear, even though in reality, it may not be. 
It aims to find the minimal perturbation, corresponding to the orthogonal projection onto the hyperplane.

%%%%%%%%%%%%%%%%%%%%%%%%%%%%%%%%%%%%%%%%%%%%%%%%%%%%%%%%%%%%%%%%%%%%%%%%%%%%
\subsection{Advantages of Post-Hoc OOD Detectors} \label{sec:posthocmethods}

Post-hoc OOD detection methods, which make use of specific layers of the pre-trained classifier, have been demonstrated to outperform retraining-based approaches, thereby underscoring their empirical efficacy \cite{zhang2023openood}.
Their plug-and-play nature allows seamless integration with pre-trained models, without necessitating alterations to the training procedure or access to original training data \cite{zhang2023openood}.
In \cref{tab:taxonomy} are 16 post-hoc detectors listed, which also shows if a method uses features, logits, or the probabilities of the pre-trained model.
Post-hoc methods underline their simplicity to outperform others on natural distribution shifts datasets by being as lightweight as possible. 
The latest post-hoc OOD detector in \cref{tab:taxonomy} is SCALE \cite{xu2023scaling}. 
In contrast to the previous activation shaping method,  ASH \cite{djurisic2022extremely}, that involves pruning and scaling of activations, SCALE demonstrates sota results are solely derived from scaling.

OOD detection has a more expansive scope when compared to anomaly detection or open-set recognition (OSR) \cite{scheirer2012toward}.
Anomaly detection is concerned with the identification of rare deviations within a single distribution.
OSR  addresses the issue of unknown classes during inference. 
OOD detection methods aim to identify any test sample that deviates from the training data distribution \cite{zhang2023openood}.

Moreover, post-hoc OOD methods can be augmented with other techniques, such as those employed in OSR \cite{gillert2021towards} or uncertainty estimation \cite{schwaiger2020uncertainty}. 
Combining different techniques means also the post-hoc methods become more complex and thus the attack surface might become larger, i.e. attacks against uncertainty estimation   \cite{ledda2023adversarial} differ from the evasion attacks.
It is not necessarily the case that a post-hoc method must be combined with other techniques.
This demonstrates SAFE \cite{wilson2023safe}.
% Post-hoc method must not necessarily combine with other techniques to become more robust against evasion attacks, as demonstrated recently by SAFE \cite{wilson2023safe}.

%%%%%%%%%%%%%%%%%%%%%%%%%%%%%%%%%%%%%%%%%%
\subsection{OOD Adversarial Detection} \label{sec:adv_defenses}
Ensuring the protection of deployed DL models is the aim of OOD detectors. 
Merely detecting \cite{tramer2022detecting} is already a challenging, but the task of providing comprehensive defense \cite{carlini2019evaluating} against unknown threats is challenging. 
Every defense mechanism can be circumvented at some point \cite{carlini2017adversarial}.
Many OOD detectors can be easily evaded by slightly perturbing benign OOD inputs, creating OOD AdEx that reveals a severe limitation of current open-world learning frameworks \cite{sehwag2019better, azizmalayeri2022your}.
Even adversarial training-based defense methods, effective against ID adversarial attacks, struggle against OOD AdEx \cite{azizmalayeri2022your}.
In the past years, various defensive techniques \cite{wu2023defenses} and combinations of them have surfaced to combat threats such as adversarial training  \cite{madry2017towards,wang2023better, bai2024mixednuts}, gradient masking for obfuscation \cite{papernot2017practical},  input transformations such as input purification \cite{nie2022diffusion, lin2024adversarial}. % certified defenses \cite{wang2024certified}, and detection \cite{harder2021spectraldefense, peng2024aed}.
However, attackers consistently adapt their adversarial attacks to the specific defense mechanisms  \cite{tramer2020adaptive}.
According to \cite{croce2022evaluating}, optimization-based defenses could be a promising future, because they can adapt during test-time towards the input.
% There has been a recent shift towards exploring defenses that adapt during test-time, as highlighted in the study by \cite{croce2022evaluating}. 
% , as these attacks exhibit a remarkable ability to transfer effectively across diverse datasets \cite{alhamoud2022generalizability} and models \cite{gu2023survey}.

OOD detectors have benefited from insights in the adversarial machine learning (AML) field, but still lack comprehensive defense against unknown threats. 
There are adversarial training based methods, e.g. ALOE \cite{chen2020robust}, OSAD \cite{shao2020open}, or  ATOM  \cite{chen2021atom}. 
A discriminator-based method, ADT \cite{azizmalayeri2022your},  significantly outperforms previous methods by addressing their vulnerabilities to strong adversarial attacks.
More recently, the post-hoc method  SAFE \cite{wilson2023safe} leverages the most sensitive layers in a pre-trained classifier through targeted input-level adversarial perturbations. 
To this end, ``adversarial robust'' OOD detection methods lag, being a comprehensive defense against unknown and adaptive threats remains an intricate challenge.

\subsection{CAM-based Explanations and  Consistent Shift by AdEx} \label{sec:releatedgradcam}
Explainable AI (XAI) \cite{longo2024explainable} can be used as a defense mechanism \cite{wu2018attention, stiff2019explainable, rieger2020simple} to spot AdEx \cite{noppel2023sok}.
The Grad-CAM heatmap provides a clear indication, for a given image, of the regions that the Convolutional Neural Network (CNN) focuses on to make a decision. 
% Areas in blue on the heatmap signify minimal involvement, while red areas indicate high participation. 
Class Activation Maps (CAMs) can be considered saliency maps specific to the input \cite{zhou2016learning}. 
These maps emerge from the accumulated and up-scaled activations at a particular convolutional layer, typically the penultimate layer. 
The classification is estimated through a linear combination of the activation of units in the final layer(s) of the feature selection network:
% \begin{equation}
% f_\theta(.) \approx \sum_i \sum_k w_k a_{ki},
% \end{equation}
$f_\theta(.) \approx \sum_i \sum_k w_k a_{ki},$
where $a_{ki}$ represents the activation of the k-th channel of unit $i$, and $w_k$ denotes the learned weights. 
The relevance values are subsequently defined as $r_i=\sum_kw_k a_{ki}$.
The determination of these weights depends on the specific CAM variant employed \cite{selvaraju2016grad, selvaraju2017grad, wang2020score}.
Grad-CAM weights the activations using gradients:
$w_K:=\frac{\partial f_\theta (.)}{\partial a_{ki}}.$
% \begin{equation}
%     w_K:=\frac{\partial f_\theta (.)}{\partial a_{ki}}.
% \end{equation}
This weighting directly links to more fundamental explanations that merely estimate the influence of the input on the final output as described before $r_i=\partial f_\theta (.) / \partial \vx_i$ \cite{binder2013enhanced}.

In context to AdEx, 
% Chakraborty et al. 
\cite{chakraborty2022generalizing} 
% introduces an innovative approach to expand Grad-CAM beyond individual examples, transforming it into a method for elucidating the overall behavior of the model.
% This novel method extends Grad-CAM from individualized explanations to a technique that explains the global behavior of the model. 
 observes a consistent shift in the region highlighted in the gradient activation heatmap, reflecting its contribution to decision-making across all models under adversarial attacks. 
% The authors address a limitation of Grad-CAM, namely its difficulty in generalizing  CNN behavior.

%%%%%%%%%%%%%%%%%%%%%%%%%%%%%%%%%%%%%%%%%%%%%%%%%%%%%%%%%%%%%%%%%%%%%%%%%%%%%%%%%%%%%%%%%%%%%%%%%%%%%%%%%%%%%
\section{DEFINITIONS IN OOD DETECTION} \label{sec:definition}
% \section{STANDARDIZING THE DEFINITION - AdEx AS OUT DISTRIBUTION} \label{sec:definition}
\begin{table*}[h]
\centering
\caption{
Overview of ID and OOD definition from several OOD detectors.
The post-hoc detectors are divided into:
 \colorbox[HTML]{FED8E5}{Classifiction-based}, \colorbox[HTML]{CFE7FF}{Density-based}, \colorbox[HTML]{F3D5B0}{Distance-based} \cite{zhang2023openood}.
 We also mark: \colorbox[HTML]{70AD47}{Supervised and Adversarial Robust}, which are not post-hoc detectors.
} \label{tab:def}
\resizebox{1\textwidth}{!}{\small
\begin{tabular}{p{0.14\textwidth}|p{0.17\textwidth}p{0.36\textwidth}p{0.33\textwidth}}
    \toprule[1pt]
    \hline
    \textbf{Methods} &
      \textbf{ID} &
      \textbf{OOD} &
      \textbf{Model Architectures} \\ \hline     \hline   
    \multicolumn{4}{c}{Post-Hoc OOD Detection Methods} \\ \hline    
    % OpenOOD \cite{zhang2023openood} &
    %   ImageNet-1K &
    %   Near-OOD: SSB-hard, NINCO; Far-OOD: iNaturalist, Textures, OpenImage-O; Covariate-Shifted ID: ImgeNet-C, ImageNet-R, ImageNet-v2 &
    %   ResNet-50, Swin-T,   VIT-B-16 \\ \hline
    \rowcolor[HTML]{FED8E5} 
    SCALE \cite{xu2023scaling} &
      ImageNet-1K &
      Near-OOD: NINCO, SSB-hard; Far-OOD: iNaturalist,  OpenImage-O, Textures
      & ResNet-50 \\
    \rowcolor[HTML]{FED8E5} 
    NNGuide \cite{park2023nearest} &
      ImageNet-1K &
      Near-OOD: iNaturalist, OpenImage-O; Far-OOD: Textures; Overlapping: SUN and Places &
      MobileNet, RegNet, ResNet-50, ViT \\
    \rowcolor[HTML]{FED8E5} 
    GEN \cite{liu2023gen} &
      ImageNet-1K &
      ImageNet-O, iNaturalist, OpenImage-O, Texture,  &
      BiT, DeiT, RepVGG, ResNet-50, ResNet-50-D, Swin-T, ViT \\
    \rowcolor[HTML]{FED8E5} 
    ASH \cite{djurisic2022extremely} &
      ImageNet-1K &
      iNaturalist, Places, SUN, Textures &
      MobileNetV2, ResNet-50 \\
    \rowcolor[HTML]{CFE7FF} 
    DICE \cite{sun2022dice} &
      ImageNet-1K &
        iNaturalist, Places, SUN, Textures &
      DenseNet-101 \\
    \rowcolor[HTML]{FED8E5} 
    KNN \cite{sun2022out} &
      ImageNet-1K &
        iNaturalist, Places, SUN, Textures &
      ResNet-50 \\
    \rowcolor[HTML]{FED8E5} 
    VIM \cite{wang2022vim} &
      ImageNet-1K &
      ImageNet-O, iNaturalist, OpenImage-O, Texture &
      BiT-S, DeiT,  RepVGG, ResNet-50, ResNet-50-D, Swin-T, VIT-B-16 \\
    \rowcolor[HTML]{FED8E5} 
    KLM; MLS \cite{hendrycks2019scaling} & ImageNet-21K; ImageNet-1K, Places & Species (categories); BDD-Anomaly, StreetHazards (segmentation)
      & Mixer-B-16; ResNet-50, TResNet-M, ViTB-16 \\
    \rowcolor[HTML]{FED8E5} 
    REACT \cite{sun2021react} &
      ImageNet-1K &
      iNaturalist, Places, SUN, Textures &
      MobileNet, ResNet \\
    % \rowcolor[HTML]{FED8E5} 
    % MLS &
    %   ImageNet-21K-P, ImageNet-1K, Places, PASCAL VOC, MS-COCO &
    %   Categories from the Species dataset, 20   out-of-distribution classes from  ImageNet-21K. These classes have no overlap with ImageNet-1K, PASCAL VOC, or MS-COCO &
    %   ResNet, ViT, MLP mixer \\
    \rowcolor[HTML]{FED8E5} 
    GRAM \cite{huang2021importance}&
      ImageNet-1K &
      iNaturalist, SUN, Places, Textures &
      DenseNet-121, ResNetv2-101 \\
    \rowcolor[HTML]{F3D5B0} 
    RMDS \cite{ren2021simple} &
       CIFAR-10, CIFAR-100 &
      CIFAR-10,   CIFAR-100 &
      BiT, CLIP, VIT-B-16 \\
    \rowcolor[HTML]{CFE7FF} 
    EBO \cite{liu2020energy} &
      CIFAR-10 &
      ISUN, Places, Texture, SVHN, LSUN &
      WideResNet \\
    \rowcolor[HTML]{F3D5B1} 
    MDS \cite{lee2018simple} &
      CIFAR-10 &
      SVHN, TinyImageNet, LSUN, AdEx &
      DenseNet, ResNet \\
    \rowcolor[HTML]{FED8E5} 
    ODIN  \cite{liang2017enhancing} &
      CIFAR-10 &
      LSUN, SVHN, TinyImageNet &
      DenseNet, ResNet \\
    \rowcolor[HTML]{FED8E5} 
    MSP \cite{hendrycks2016baseline} &
      CIFAR-10 &
      SUN (Gaussian) &
      WideResNet 40-4 \\ 
    \hline
    \multicolumn{4}{c}{ Adversarial Robust OOD Detectors} \\ \hline    
    \rowcolor[HTML]{70AD47} 
    ALOE \cite{chen2020robust} &
      CIFAR-10, CIFAR-100, GSTRB &
      PGD attack &
      DenseNet \\
    \rowcolor[HTML]{70AD47} 
    OSAD \cite{shao2020open}  &
      CIFAR-10, SVHN, TinyImageNet &
      FGSM, PGD attack &
      ResNet-18 \\
    \rowcolor[HTML]{70AD47} 
    ADT \cite{azizmalayeri2022your} &
      CIFAR-10, CIFAR-100 &
      FGSM, PGD attack &
      ViT \\
    \rowcolor[HTML]{70AD47} 
    ATOM \cite{chen2021atom} &
       CIFAR-10, CIFAR-100, SVHN &
      PGD attack &
      WideResNet \\
    % \rowcolor[HTML]{70AD47} 
    % HAT \cite{rade2021reducing} &
    %    CIFAR-10, CIFAR-100, SVHN &
    %   PGD attack&
    %   WideResNet \\
    \rowcolor[HTML]{FED8E5} 
    SAFE \cite{wilson2023safe} &
      PASCAL-VOC, DeepDrive &
      FGSM attack &
      RegNetX4.0, ResNet-50 \\
    \hline \bottomrule[1pt]
\end{tabular}
}
\end{table*}
The robustness definition in the field of OOD detectors has been ambiguous when it comes to attack methods. 
%Additionally, the performance of attacks methods rely on  hyperparameters selection, models, and datasets. 
There are two categories of AdEx.  
The first one merely attacks the underlying pre-trained classifier and the second one aims to fool the OOD detector itself.
Adversarial robustness can be considered in a classifier (Unified Robustness) or the OOD detector (Robust OOD Detection) according to \cite{karunanayake2024out}.
% In a recent survey , the authors defined this as ``Unified Robustness'' and  ``Robust OOD Detection''  respectively.
This work focuses on unified robustness, which belongs to the covariate shift.
For image classification, a dataset $\train = \{(\vx_i, y_i); \vx_i \in \gX, y_i \in \gY\}$ sample from a training distribution $\Ptrain(\vx, y)$ is used to train some classifier $C: \gX \rightarrow \gY$. 
In real-world deployments, distribution shift occurs when classifier $C$ receives data from test distribution $\Ptest(\vx,y)$ where $\Ptrain(\vx, y) \neq \Ptest(\vx,y)$ \cite{moreno2012unifying}.
An OOD detector is a scoring function $s$ that maps an image $\vx$ to a real number $\R$ such that some threshold $\tau$ arrives at the detection rule $f(\vx): \text{ID if } s(\vx) \geq \tau \text{, OOD otherwise}$.
% Common changes within the data distribution fall under two categories: covariate shift, which is label-preserving (i.e. concerns examples only from training classes), and semantic shift, which is label-altering (concerns examples only from new classes).
\paragraph{Existing Definitions}
We present a comparison of the selected ID and OOD datasets utilized by each OOD detector from \cref{tab:taxonomy} in  \cref{tab:def}. %in \cref{app:definition}. 
Our focus lies on experiments employing the most extensive available datasets. 
Additionally, we enrich this analysis by including adversarial robust OOD detectors discussed in 
% section 2.3. 
\cref{sec:adv_defenses}. 
Notably, recent post-hoc detectors predominantly utilize ImageNet-1K as their standard ID datasets, whereas adversarial robust OOD detectors are more commonly evaluated on the smaller CIFAR-10 dataset. 
Furthermore, the architecture of adversarial robust methods diverges from lightweight post-hoc designs, often relying on adversarial training and auxiliary data, as elaborated in 
% section 2.3. 
\cref{sec:adv_defenses}.
The popular adversarial ODD are the FGSM and PGD attacks. 
The \cref{tab:def} gives an overview of several detectors as well as considered ID and OOD datasets and model architectures.
The ImageNet-1K dataset together with  ResNet-50  architecture has become standard for ID.
Popular OD datasets are iNaturalist, SUN, Places, and Textures.
Some OOD detectors, such as ALOE, OSAD, ADT, and ATOM (see 
\cref{sec:adv_defenses}), aim to be robust adversarial. 
They usually take evading attacks (see 
\cref{sec:evasionattacks}) such as FGSM or PGD as OOD.
These computationally expensive methods only show empirical results on the small-scale CIFAR-10.
% The authors from ALOE \cite{chen2020robust} and ADT \cite{azizmalayeri2022your}  even explicitly write that evasion attacks do not change the semantic. 
A robust OOD detector is built to distinguish whether a perturbed input is OOD.
Standardized OOD  benchmark frameworks, i.e.  OpenOOD \cite{zhang2023openood} or RoboDepth \cite{kong2024robodepth} do not currently include unified robustness in their benchmarks.
Consequently, both frameworks, give a false sense of encompassing open-world capabilities. 
They focus on natural distribution \cite{hendrycks2021natural}, where OOD detection in large-scale semantic space has attracted increasing attention \cite{hendrycks2019scaling}. 
Some OOD datasets have issues, where ID classes are part of the OD dataset \cite{bitterwolf2023or}.
Recently, \cite{yang2023imagenet} found a clean semantic shift dataset that minimizes the interference of covariate shift.
The experiments show that sota OOD detectors are more sensitive to covariate shifts, and the advances in semantic shift detection are minimal. 

\paragraph{Extension to Adversarial Robust Definition}
Investigating AdEx could gain insights into understanding the covariate shift and towards a generalized OOD detection \cite{yang2021generalized}.
The difference between benign $\vx$ and the attacked counterpart $\vx'$ is the different attention of the pre-trained classifier $C$ per sample.
Let us define an attention map $A$ \cite{guo2022attention}: The input image is passed through a classifier $C$  to obtain a feature map $\mF$. 
A possible tool could be Grad-CAM \cite{selvaraju2017grad} (see also \cref{sec:releatedgradcam}) to investigate the attention change between benign $\vx$  and adversarial example $\vx'$ on $\mF$ \cite{rieger2020simple} (compare 
\cref{sec:flmetric}).

Let $\Omega(\vx)$ be a set of semantic-preserving perturbations on an input $\vx$. 
For $ \boldsymbol\delta\in \Omega(\vx)$, $\vx + \boldsymbol\delta$ have the same distributional membership (i.e., $\vx$ and $\vx+\boldsymbol\delta$ both belong to ID or OOD) \cite{chen2020robust}.
To clarify, the attention maps $\mA$ between benign $\vx$  and  adversarial example $\vx'$ are not the same 
\begin{equation} \label{eq:attention}
    \mA_C(\vx) \neq \mA_C(\vx')\text{.} 
\end{equation}
So AdEx are not from a completely different distribution, but rather adversarial perturbed versions of ID data.

\section{EXPERIMENTS}
In this section, we explain our experimental setup and discuss the results of 16 post-hoc OOD detectors on different datasets and pre-trained models.

\subsection{Experiment Setup} %\label{sec:experimentsetup}
We extend the OpenOOD framework \cite{zhang2023openood} to consider adversarial attacks. 
We attack the pre-trained classifiers on the corresponding test sets and evaluate 16 post-hoc OOD detectors.

As \textbf{attack methods}, we choose FGSM(-$L^{\infty}$), PGD(-$L^{\infty}$),  DF(-$L^2$) from FoolBox \cite{rauber2017foolbox}; and  the mPGD-($L^{\infty}$).
The attacked \textbf{models} are ResNet-18 \cite{he2016deep}, ResNet-50 \cite{he2016deep}, and Swin-T \cite{liu2021swin}. 
The \textbf{datasets} are  CIFAR-10 \& CIFAR-100 \cite{krizhevsky2009cifar}, ImageNet-1K \cite{deng2009imagenet} and its variant ImageNet-200 with just 200 classes.
\begin{table}[ht]
\centering
\caption{
Setup. The attack success rate (ASR) from various attacks on different models and datasets.
The FGSM attack has the lowest, while DF has the highest ASR.
$\mathcal{A}_{\text{std}}$ refers to the standard accuracy of the pre-trained classifier.
} \label{tab:asr}
\resizebox{\linewidth}{!}{
    \begin{tabular}{l|lr|lr}
    \toprule[1pt]
    \hline
     \textbf{Dataset} & \textbf{Arch} & $\mathcal{A}_{\text{std}}$ \textbf{(\%)} & \textbf{Attack} & \textbf{ASR (\%)} \\ \hline
        CIFAR-10     & ResNet-18 & 95.32 & PGD  & 99.88 \\
                     &          &       & FGSM & 59.21 \\
                     &          &       & DF   & 100   \\
                     &          &       & mPGD & 68.06 \\ \hline
        CIFAR-100    & ResNet-18 & 77.19 & PGD  & 100   \\
                     &          &       & FGSM & 91.99 \\
                     &          &       & DF   & 100   \\
                     &          &       & mPGD & 88.14 \\ \hline
        ImageNet-200 & ResNet-18 & 86.27 & PGD  & 99.9  \\
                     &          &       & FGSM & 95.46 \\
                     &          &       & DF   & 100   \\
                     &          &       & mPGD & 96.53 \\ \hline
        ImageNet-1K  & ResNet-50 & 76.19 & PGD  & 99.97 \\
                     &          &  & FGSM & 93.33 \\
                     &          &  & DF   & 100   \\
                     &          &  & mPGD & 98.48 \\
                     & Swin-T   & 95.99 & PGD  & 99.99 \\
                     &          &  & FGSM & 75.09 \\
                     &          &  & DF   & 100   \\
                     &          &  & mPGD & 98.84 \\ 
                     % & ViT-B-16 & PGD    & 98.44    \\
                     % &          & FGSM   & 69.95    \\
                     % &          & DF     & 99.99    \\
                     % &          & mPGD   & 96.28   \\ 
                 \hline \bottomrule[1pt]
    \end{tabular}
    }
\end{table}
% The attacks are not always successful and depend on the model architecture and dataset and list the attack success rate (ASR) in \cref{tab:asr}.
The efficacy of the attacks is not absolute and depends on a multitude of factors, including the hyperparameters, model architecture, and the dataset. 
The attack success rate (ASR) is presented in \cref{tab:asr}. 
% It must be considered that the underlying attacked test sets strength depends on the ASR of each attack method, comparing \cref{tab:asr} with  \cref{tab:results}.
The attacks PGD and FGSM do have an epsilon size, we use an epsilon size of $8/255$ for CIFAR-10/100 and $4/255$ for the ImageNet. 
The mPGD randomly attacks an area of the image ($8 \times 8$ px for CIFAR-10/100 and  $60 \times 60$ px for the ImageNet) without an epsilon constraint, leading to perceptible perturbations.

We utilize two \textbf{metrics} to assess the OOD detection performance, elaborated as follows:
{1) FPR95$\downarrow$} stands for false positive rate measured when true positive rate (TPR) sits at 95\%.
Intuitively, FPR95 measures the portion of samples that are falsely recognized as ID data.
{2) AUROC$\uparrow$} refers to the area under the receiver operating characteristic curve for binary classification problems like  OOD detection.

\begin{figure*}[h] 
    \centering
    \includegraphics[width=1\textwidth]{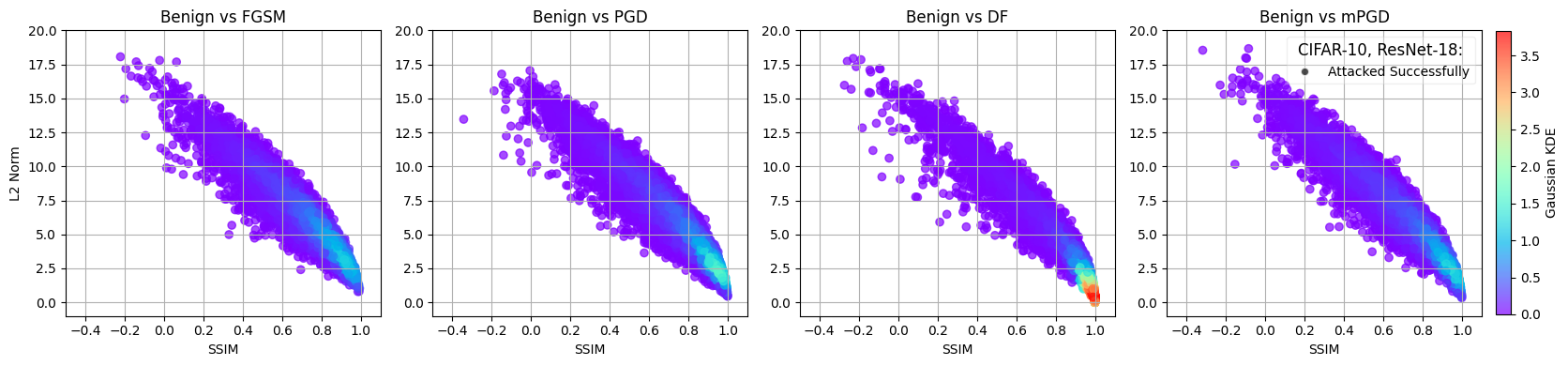}
    \includegraphics[width=1\textwidth]{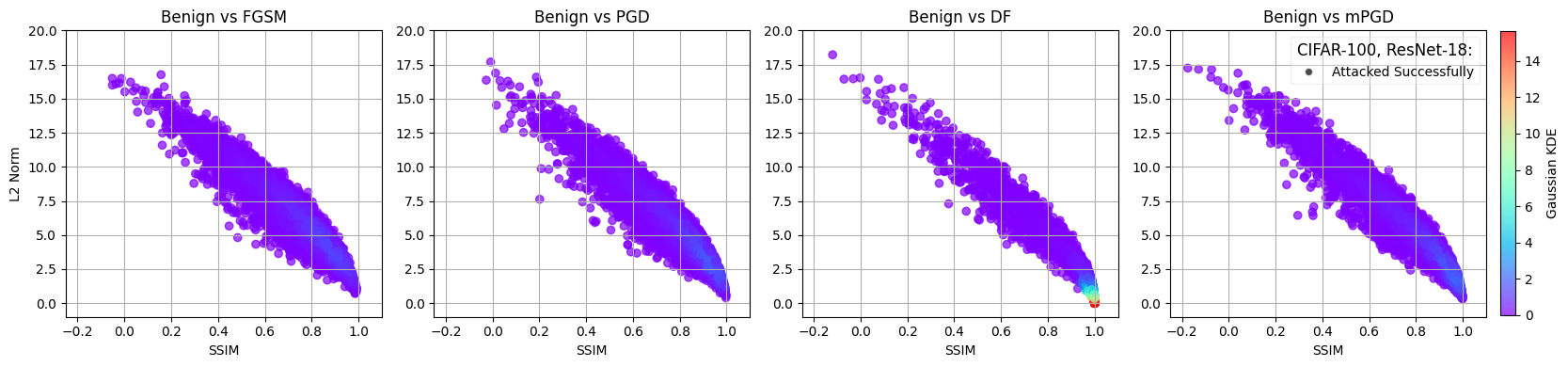}
    \includegraphics[width=1\textwidth]{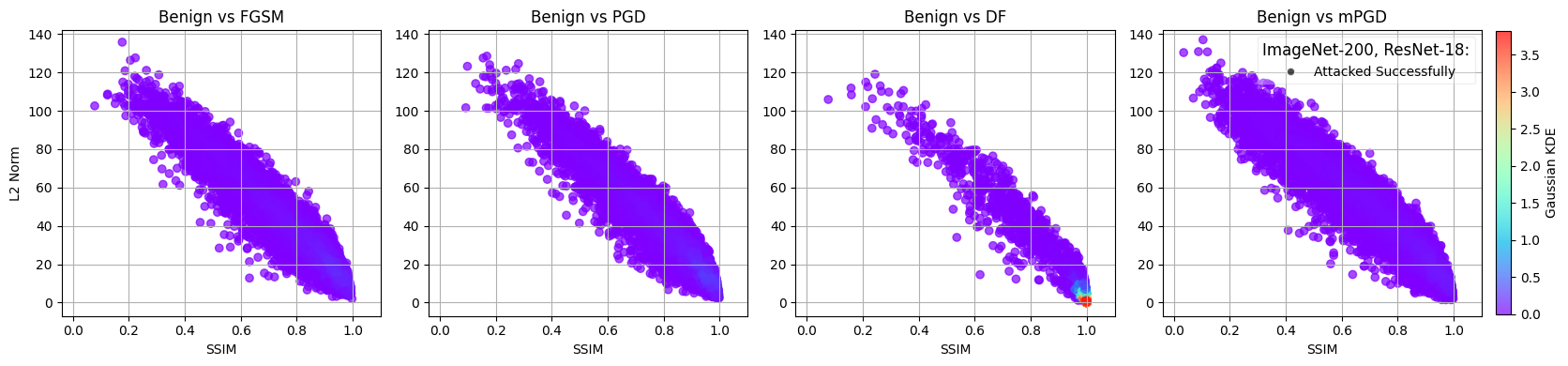}
    \includegraphics[width=1\textwidth]{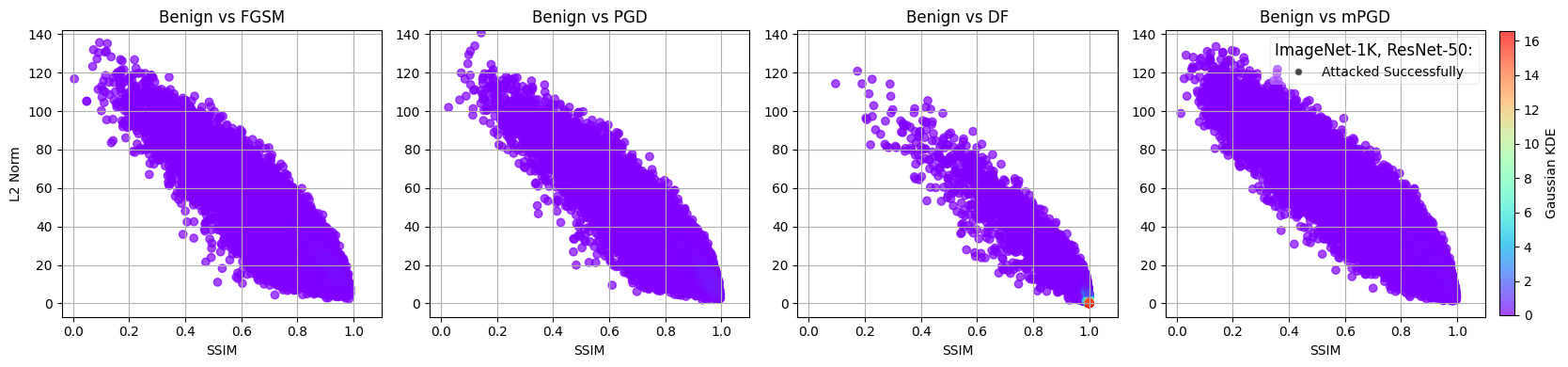}
    \includegraphics[width=0.99\textwidth]{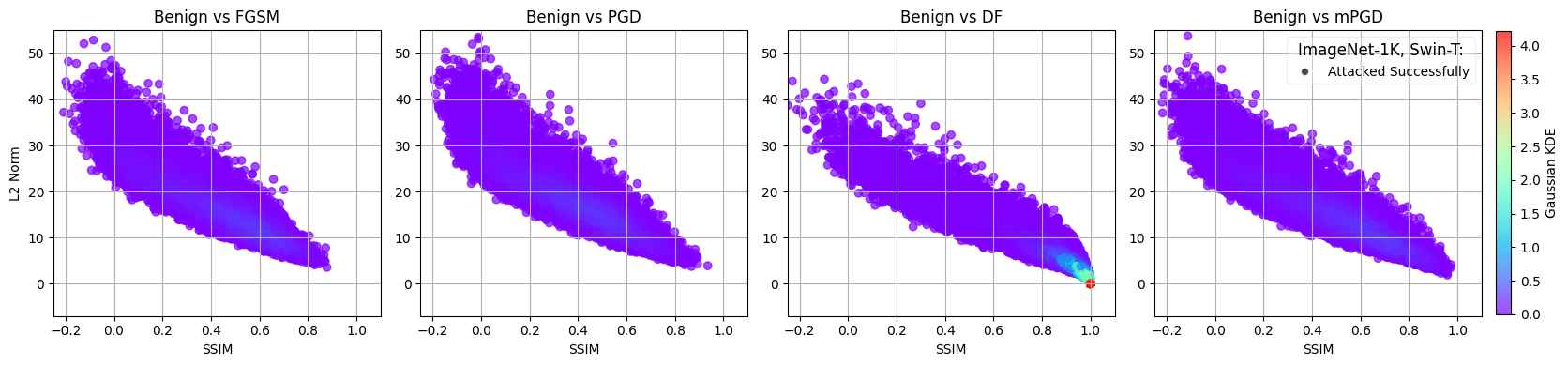}    
    \caption{
        Grad-CAM comparison between the benign and its attacked counterpart.
        The color red indicates a high intensity of samples on similar heatmaps. 
        The color blue indicates a low intensity.
        The attacks can be compared row-wise that is sorted according to datasets and attacked DNN.
        The DF attack yields very similar heatmaps across all datasets and models.
    } 
    \label{fig:gradcam}
\end{figure*}

\begin{table*}[!h]
\caption{
Results. 
We evaluate the post-hoc OOD detectors using the metrics FPR95$\downarrow$ (\%) and AUROC$\uparrow$ (\%). 
The norm-bounded attacks PGD and FGSM do have an epsilon size of $8/255$ for CIFAR-10/100 and $4/255$ for the ImageNet. 
} \label{tab:results}
\centering
% \resizebox{0.85\linewidth}{!}{%
\begin{adjustbox}{width=1.0\linewidth}
    \begin{tabular}{ll||rr:rr:rr:rr:rr}
    \toprule[1pt]
    \hline
    \multirow{3}{*}{\textbf{Detector}} & \multirow{3}{*}{\textbf{Attacks}} & \multicolumn{2}{c}{\textbf{CIFAR-10}} & \multicolumn{2}{c}{\textbf{CIFAR-100}} & \multicolumn{2}{c|}{\textbf{ImageNet-200}} & \multicolumn{4}{c}{\textbf{ImageNet-1K}} \\
     &  & \multicolumn{6}{c|}{\textbf{ResNet-18}} & \multicolumn{2}{c}{\textbf{ResNet-50}} & \multicolumn{2}{c}{\textbf{Swin-T}} \\
     &  & \textbf{FPR95} & \textbf{AUROC} & \textbf{FPR95} & \textbf{AUROC} & \textbf{FPR95}& \multicolumn{1}{r|}{\textbf{AUROC}} & \textbf{FPR95} & \textbf{AUROC} & \textbf{FPR95} & \textbf{AUROC} \\ 
    \hline \hline
    \multirow{4}{*}{SCALE} & PGD & 99.67 & 34.53 & 99.97 & 16.18 & 95.49 & 35.14 & 100.00 & 0.20 & 95.49 & 35.14 \\
     & FGSM & 85.74 & 77.50 & 49.69 & 85.64 & 79.75 & 76.88 & 89.75 & 66.28 & 79.75 & 76.88 \\
     & DF & 67.07 & 81.73 & 69.22 & 68.69 & 79.75 & 76.88 & 87.82 & 57.77 & 79.75 & 76.88 \\
     & mPGD & 88.50 & 70.69 & 85.58 & 59.67 & 93.24 & 42.17 & 100.00 & 6.90 & 93.24 & 42.17 \\
    \hline
    \multirow{4}{*}{NNGUIDE} & PGD & 99.39 & 30.29 & 98.85 & 17.07 & 96.44 & 33.60 & 100.00 & 0.12 & 96.44 & 33.60 \\
     & FGSM & 93.10 & 53.01 & 68.14 & 77.62 & 83.21 & 75.13 & 85.27 & 73.53 & 83.21 & 75.13 \\
     & DF & 92.08 & 63.25 & 85.36 & 64.14 & 83.21 & 75.13 & 82.19 & 62.81 & 83.21 & 75.13 \\
     & mPGD & 92.94 & 58.90 & 90.98 & 57.26 & 94.30 & 42.87 & 99.99 & 9.52 & 94.30 & 42.87 \\ 
    \hline
    \multirow{4}{*}{GEN} & PGD & 99.51 & 41.75 & 99.90 & 26.03 & 89.17 & 40.03 & 100.00 & 0.21 & 89.17 & 40.03 \\
     & FGSM & 70.14 & 81.29 & 45.66 & 87.10 & 72.06 & 79.00 & 83.63 & 73.28 & 72.06 & 79.00 \\
     & DF & 44.32 & 85.98 & 71.38 & 65.82 & 72.06 & 79.00 & 80.01 & 62.96 & 72.06 & 79.00 \\
     & mPGD & 83.65 & 74.35 & 75.59 & 66.56 & 88.10 & 47.17 & 99.96 & 12.40 & 88.10 & 47.17 \\ 
    \hline
    \multirow{4}{*}{ASH} & PGD & 99.67 & 31.23 & 99.96 & 24.49 & 97.06 & 32.97 & 100.00 & 0.19 & 97.06 & 32.97 \\
     & FGSM & 86.94 & 70.60 & 42.86 & 88.14 & 83.81 & 74.57 & 85.61 & 69.99 & 83.81 & 74.57 \\
     & DF & 77.10 & 74.62 & 74.98 & 65.44 & 83.81 & 74.57 & 83.29 & 60.88 & 83.81 & 74.57 \\
     & mPGD & 90.75 & 64.35 & 78.87 & 66.61 & 95.20 & 40.17 & 100.00 & 8.49 & 95.20 & 40.17 \\ 
    \hline
    \multirow{4}{*}{DICE} & PGD & 96.65 & 36.09 & 99.93 & 23.45 & 95.03 & 34.24 & 100.00 & 0.11 & 95.03 & 34.24 \\
     & FGSM & 75.86 & 72.65 & 46.27 & 86.51 & 75.62 & 78.73 & 86.99 & 71.20 & 75.62 & 78.73 \\
     & DF & 68.84 & 73.44 & 76.73 & 65.46 & 75.62 & 78.73 & 84.72 & 62.92 & 75.62 & 78.73 \\
     & mPGD & 89.46 & 66.25 & 80.48 & 66.68 & 91.60 & 42.99 & 99.99 & 8.63 & 91.60 & 42.99 \\ 
    \hline
    \multirow{4}{*}{KNN} & PGD & 64.91 & 69.18 & 90.06 & 43.07 & 85.23 & 55.53 & 78.63 & 55.74 & 85.23 & 55.53 \\
     & FGSM & 61.23 & 82.08 & 47.81 & 84.69 & 76.88 & 73.23 & 75.89 & 68.43 & 76.88 & 73.23 \\
     & DF & 38.78 & 85.84 & 78.54 & 63.06 & 76.88 & 73.23 & 86.65 & 58.09 & 76.88 & 73.23 \\
     & mPGD & 76.02 & 75.02 & 78.93 & 64.03 & 88.63 & 54.44 & 87.09 & 48.40 & 88.63 & 54.44 \\ 
    \hline
    \multirow{4}{*}{VIM} & PGD & 92.45 & 56.83 & 98.16 & 42.79 & 89.17 & 46.44 & 100.00 & 5.16 & 89.17 & 46.44 \\
     & FGSM & 54.89 & 84.43 & 54.72 & 74.75 & 71.60 & 69.44 & 75.28 & 71.31 & 71.60 & 69.44 \\
     & DF & 43.92 & 84.92 & 78.57 & 61.75 & 71.60 & 69.44 & 82.95 & 59.88 & 71.60 & 69.44 \\
     & mPGD & 80.70 & 74.35 & 80.56 & 60.01 & 90.16 & 50.71 & 99.86 & 24.51 & 90.16 & 50.71 \\ 
    \hline
    \multirow{4}{*}{KLM} & PGD & 91.43 & 60.91 & 91.94 & 45.75 & 90.87 & 54.58 & 95.49 & 40.77 & 90.87 & 54.58 \\
     & FGSM & 96.90 & 66.02 & 72.31 & 80.83 & 80.71 & 74.44 & 80.54 & 71.56 & 80.71 & 74.44 \\
     & DF & 80.84 & 71.64 & 91.38 & 59.58 & 80.71 & 74.44 & 85.48 & 59.30 & 80.71 & 74.44 \\
     & mPGD & 97.52 & 62.26 & 89.69 & 61.30 & 91.96 & 55.14 & 94.81 & 41.19 & 91.96 & 55.14 \\ 
    \hline
    \multirow{4}{*}{MLS} & PGD & 99.58 & 39.65 & 99.96 & 24.43 & 94.43 & 35.64 & 100.00 & 0.12 & 94.43 & 35.64 \\
     & FGSM & 75.61 & 80.97 & 43.04 & 87.63 & 74.47 & 79.06 & 85.30 & 74.16 & 74.47 & 79.06 \\
     & DF & 51.12 & 84.89 & 74.91 & 65.41 & 74.47 & 79.06 & 81.44 & 63.37 & 74.47 & 79.06 \\
     & mPGD & 85.11 & 73.78 & 78.81 & 66.41 & 90.90 & 44.19 & 99.97 & 10.65 & 90.90 & 44.19 \\ 
    \hline
    \multirow{4}{*}{REACT} & PGD & 98.84 & 45.19 & 99.89 & 25.13 & 94.49 & 35.90 & 100.00 & 4.15 & 94.49 & 35.90 \\
     & FGSM & 79.55 & 79.84 & 42.83 & 88.27 & 76.12 & 78.15 & 80.31 & 74.14 & 76.12 & 78.15 \\
     & DF & 54.80 & 84.16 & 74.99 & 65.49 & 76.12 & 78.15 & 79.97 & 62.88 & 76.12 & 78.15 \\
     & mPGD & 85.32 & 73.19 & 78.70 & 66.52 & 91.26 & 44.59 & 99.71 & 20.13 & 91.26 & 44.59 \\ 
    \hline
    \multirow{4}{*}{GRAM} & PGD & 99.82 & 22.50 & 99.94 & 17.12 & 98.75 & 25.92 & 100.00 & 0.07 & 98.75 & 25.92 \\
     & FGSM & 94.77 & 56.54 & 79.42 & 79.23 & 91.60 & 69.07 & 96.06 & 59.67 & 91.60 & 69.07 \\
     & DF & 88.14 & 60.87 & 90.78 & 58.04 & 91.60 & 69.07 & 93.66 & 55.04 & 91.60 & 69.07 \\
     & mPGD & 93.21 & 55.56 & 91.49 & 58.72 & 97.83 & 34.92 & 100.00 & 4.36 & 97.83 & 34.92 \\ 
    \hline
    \multirow{4}{*}{RMDS} & PGD & 49.03 & 82.70 & 66.28 & 77.08 & 53.69 & 76.40 & \textbf{37.47} & \textbf{95.23} & 53.69 & 76.40 \\
     & FGSM & 68.03 & 80.66 & 76.00 & 79.46 & 67.10 & 76.26 & 71.65 & 73.33 & 67.10 & 76.26 \\
     & DF & 43.26 & 85.75 & 64.52 & 80.96 & 67.10 & 76.26 & 73.74 & 65.11 & 67.10 & 76.26 \\
     & mPGD & 77.05 & 75.84 & 84.46 & 74.74 & 79.37 & 63.99 & 90.78 & 50.99 & 79.37 & 63.99 \\ 
    \hline
    \multirow{4}{*}{EBO} & PGD & 99.58 & 39.61 & 99.96 & 24.49 & 94.47 & 35.41 & 100.00 & 0.12 & 94.47 & 35.41 \\
     & FGSM & 75.62 & 81.04 & 42.86 & 88.14 & 74.58 & 79.17 & 85.35 & 74.42 & 74.58 & 79.17 \\
     & DF & 51.23 & 84.71 & 74.99 & 65.44 & 74.58 & 79.17 & 81.63 & 63.79 & 74.58 & 79.17 \\
     & mPGD & 85.11 & 73.80 & 78.87 & 66.61 & 91.00 & 44.10 & 99.97 & 10.64 & 91.00 & 44.10 \\ 
    \hline
    \multirow{4}{*}{MDS} 
     & PGD & 47.81 & \textbf{84.48} & 50.82 & \textbf{84.18} & 57.34 & 76.11 & \textbf{0.05} & \textbf{99.95} & 57.34 & 76.11 \\
     & FGSM & 64.24 & 79.22 & 86.06 & 51.31 & 91.41 & 49.61 & 90.61 & 52.93 & 91.41 & 49.61 \\
     & DF & 54.04 & 79.60 & 89.76 & 52.53 & 91.41 & 49.61 & 93.19 & 49.18 & 91.41 & 49.61 \\
     & mPGD & 78.25 & 69.86 & 91.87 & 50.63 & 83.81 & 65.37 & 51.74 & \textbf{89.98} & 83.81 & 65.37 \\ 
    \hline
    \multirow{4}{*}{ODIN} 
     & PGD & 99.73 & 33.25 & 99.97 & 17.58 & 97.20 & 31.19 & 100.00 & 0.91 & 97.20 & 31.19 \\
     & FGSM & 73.56 & 83.16 & 39.74 & 90.19 & 77.39 & 76.38 & 85.44 & 72.56 & 77.39 & 76.38 \\
     & DF & 60.16 & 83.99 & 72.94 & 68.00 & 77.39 & 76.38 & 82.00 & 65.46 & 77.39 & 76.38 \\
     & mPGD & 86.49 & 74.11 & 85.66 & 62.09 & 91.72 & 45.12 & 99.90 & 15.89 & 91.72 & 45.12 \\ 
    \hline
    \multirow{4}{*}{MSP} 
     & PGD & 99.34 & 43.62 & 100.00 & 27.04 & 87.66 & 42.60 & 100.00 & 4.08 & 87.66 & 42.60 \\
     & FGSM & 67.16 & 81.14 & 47.87 & 85.08 & 72.12 & 77.77 & 80.39 & 70.45 & 72.12 & 77.77 \\
     & DF & 39.78 & 86.85 & 70.63 & 65.92 & 72.12 & 77.77 & 72.65 & 60.76 & 72.12 & 77.77 \\
     & mPGD & 82.76 & 74.68 & 75.03 & 66.23 & 87.48 & 49.07 & 100.00 & 22.21 & 87.48 & 49.07 \\ 
    \hline
    \bottomrule[1pt]
    \end{tabular}
% }
\end{adjustbox}
\end{table*}

\subsection{Grad-CAM Similarity} \label{sec:flmetric}
% \mario{
This section investigates the neural networks' attention variations to classify benign and adversarial images.
Our goal is to measure if adversarial attacks produce a change in the attention of the neural network that leads to misclassification as depicted in \cref{fig:gradcam}. 
A previous study showed that misclassification is correlated with a shift in the center of attention for the of image classification \cite{chakraborty2022generalizing}. 
We want to test this attention change in our pair of benign/attacked images. 
To quantitative measure the aforementioned attention change we use the Grad-CAM method \cite{selvaraju2016grad}.
The Grad-CAM method allows us to map the relevant part of a given image that leads to the network classification. 
Usually, the last block of the model is used to obtain attention \cite{jacobgilpytorchcam}. 
In our work, we investigate the Grad-CAM output of benign and attacked images.
To quantitatively analyze the Grad-CAM difference of both samples we consider two well-known metrics: the mean square error distance (L2), and the structural similar index method (SSIM) \cite{wang2004image}. 
The L2 distance allows us to evaluate the pixel-wise difference of the Grad-CAM of both images. 
The higher the L2 value, the more distinctive the overall pixel-wise attention of the network between the benign and attacked image. 
The SSIM metric considers the overall structural similarity between the Grad-CAM of both images. 
The higher the metric the more similar the overall structure of the attention visualized through Grad-CAM maps. 
The combination of both metrics for our quantitative analyses is convenient because it allows us to understand better how/if an adversarial attack changes the attention of the network in an image for its classification: 
\begin{itemize}[nosep]
    \item High L2 and low SSIM report total dissimilarity of the Grad-CAM maps of the benign and attacked image. 
    Reporting that the Grad-CAM of the attacked image is different from the benign images.
    \item High L2 and high SSIM report pixel-wise differences of the Grad-CAM, but overall structural similarity. 
    This suggests that after the attack, the attention maps shift to a different area of the image, as noted in \cite{chakraborty2022generalizing}.
    \item Low L2 and high SSIM report total similarity, structural and pixel-wise. 
    Reporting that the Grad-CAM is not sensitive to the changes caused by the adversarial attack. 
\end{itemize}
When the L2 value approaches zero, the SSIM invariably reaches one, eliminating the possibility of simultaneously having low L2 and low SSIM.
In the \cref{fig:gradcam}, the DF attack shows a very high Grad-CAM similarity across all attack methods. 
Despite that, there is still a high dissimilarity by a high number of samples.

% \FloatBarrier
\subsection{DISCUSSION}
All 16 post-hoc methods evaluated in our study as shown in \cref{tab:results} demonstrate inadequate performance. 
Notably, only two methods based on Mahalanobis distance \cite{lee2018simple, ren2021simple} exhibit partial detection capabilities against FGSM, PGD, and mPGD attacks on ResNet-50 with ImageNet-1K. The robustness of Mahalanobis distance has been extensively explored \cite{kamoi2020mahalanobis, eustratiadis2021weight, yang2022toward, anthony2023use}, attributing its resilience to adversarial attacks to its consideration of covariance structures \cite{eustratiadis2021weight}.

Interestingly, several OOD benchmarks \cite{yang2204full, yang2022openood, zhang2023openood, kong2024robodepth} and a recent challenge \cite{kong2024robodrive} focus solely on natural distribution shifts, neglecting AdEx. 
This underscores a fundamental conflict between adversarial machine learning (AML) and  OOD detection, where detectors excel either in adversarial or natural distributions, but rarely both. 
Current OOD detectors struggle to converge on both fronts.

AdEx, as counterparts of In-Distribution (ID) data, significantly alter the attention of pre-trained classifiers, as shown in \cref{fig:gradcam} and  \cref{eq:attention}. These predominantly untargeted attacks aim to minimally change class predictions, resulting in dissimilarity in attention maps (\cref{sec:flmetric}). 
The DF attack exhibits minimal attention shifts across numerous samples, yet many attacked samples still demonstrate notable deviations in attention maps. 
Similarly, the mPGD attack induces perceptible perturbations, with slightly improved detection results compared to PGD, but overall performance remains insufficient.

Post-hoc detectors are trained on distributions similar to the underlying pre-trained model, yet studies, e.g. \cite{dong2020adversarial, lorenz2024adversarial},  indicate significant discrepancies between the training distribution and adversarial distributions. Achieving high detection rates on adversarial OOD samples is crucial for developing robust defense mechanisms, a goal that remains elusive in current research.

%%%%%%%%%%%%%%%%%%%%%%%%%%%%%%%%%%%%%%%%%%%%%%%%%%%%%%%%%%%%%%%%%%%%%%%%%%
\noindent\paragraph{Levels of Adversarial Robustness - From Detector towards Defense}
% Detection focuses on filtering out harmful data.
A defense is more sophisticated than detectors to mitigate attackers' efforts to fool a classifier.
A step towards adversarial defense could be to improve adversarial robustness in OOD detectors.
We suggest a possible roadmap to evaluate detectors and lift them toward an adversarial defense on different levels:
\begin{enumerate}[nosep, leftmargin=1.3cm, label=Level \arabic*:]
    \item Evaluate on strong attacks \cite{carlini2017adversarial} and avoid hyperparameters that weaken the strength of the attack's effect.
         The FGSM is not recommended because it performs a single step to find the adversarial perturbation, making it less effective than PGD \cite{li2020towards}.
         Furthermore, the attack hyperparameter space is huge \cite{cina2024attackbench} and could mitigate the attacks' strength.
    \item Use different models and datasets than the simple ResNet-18 trained on CIFAR-10. 
         % Many adversarial robust methods still advertise their real-world capabilities, and CIFAR-10 might be too simple.
         We suggest using ImageNet-1K because its resolution and objects are more closely related to real-world scenarios.
         % Evaluating adversarial robustness at least on ImageNet provides a more realistic assessment of a model's performance in practical applications.
    \item Elaborate your strategy to countermeasure the attack as demonstrated in \cite{sehwag2019better}. 
         New defense mechanisms have often been broken quickly again \cite{carlini2017adversarial}.
         For example, a differentiable OOD detector can be easily fooled if the attacker approximates the gradients of the network during the backward pass in a differentiable manner, known as BPDA \cite{athalye2018obfuscated}.
         % The Backward Pass Differentiable Approximation \cite{athalye2018obfuscated} is a technique to approximate the
    \item Let your method fail against sophisticated attacks, such as adaptive attacks \cite{athalye2018obfuscated, croce2022evaluating} or design OOD AdEx that convert OD to ID samples \cite{sehwag2019analyzing}. 
        Adversarial robustness is an iterative process, where defenses are proposed, evaluated, and improved upon in response to new attacks or discovered vulnerabilities. 
        % By embracing failures, continuous refinement leads to more robust methods.
\end{enumerate}
% Note that point 1 and 2 also applies for detection. 
% Point 3 and 4 on top advances a detector towards a defense. 
It must be noted that not all levels are required to be evaluated. 
This roadmap should indicate which level of OOD detection is robust. 
Further research would be encouraged to do at least an adversarial robustness benchmark (level 1), even though the OOD detector has a different focus.

\FloatBarrier
\section{CONCLUSION}
In this study, we assess the performance of the 16 post-hoc OOD detectors in their ability to detect various evasive attacks. 
We conducted white-box adversarial attacks, such as PGD and DF, on the CIFAR-10 and ImageNet-1K datasets. 
Our discovery indicates that current post-hoc methods are not ready for real-world applications as long as they are vulnerable to the well-known threat — adversarial examples (AdEx).
The semantic shift was evaluated by utilizing Grad-CAM to develop a novel metric. This metric indicates that the AdEx tends to perform a minimal shift in semantics that is spatially proximate to the original, but can exhibit a wide range across the image. 

We hope our experiments and datasets give a common baseline for further research by improving post-hoc methods towards robustness and will find a place as a standardized benchmark, such as OpenOOD. 
We hope that future post-hoc OOD detectors will include at least one level 1 evaluation, even though these detectors do not prioritize adversarial robustness.

\paragraph{Future Work}
We propose to extend the experiments towards transferability because AdEx transfers effectively across different datasets  \cite{alhamoud2022generalizability} and models \cite{mahmood2021robustness, gu2023survey}.

Then, we would suggest using black-box attacks for a more realistic open-world scenario \cite{liu2016delving, papernot2017practical}. 
Their attack patterns are usually easier to detect, but also transfer well to other models and datasets \cite{pang2018towards}.

Throughout this work, we assumed a perfect pre-trained classifier on a clean curated dataset. 
An imperfect pre-trained classifier would be the subject of future research \cite{humblot2024noisy}.